\documentclass[aps,prl,two column,floatfix,showpacs,groupedaddress,longbibliography]{revtex4}

\usepackage{graphicx}
\usepackage{hyperref}

\newcommand{\tabref}[1]{Tab.~\ref{#1}}                  

\renewcommand{\b}{\beta}

\renewcommand{\Xi}{\Xi}

\graphicspath{{./}{./figures/}}
\begin{document}

        \title{New limits on Anomalous Spin-Spin Interactions}
        \author{Attaallah Almasi}
        \altaffiliation{Present address: SRON, Netherlands Institute for Space Research}
        \email{a.almasi@sron.nl}
        \author{Junyi Lee}
        \author{Himawan Winarto}
        \author{Marc Smiciklas}
       \author{Michael V. Romalis}
        \affiliation{Department of Physics, Princeton University, Princeton, NJ 08544, USA}

        %
        %
        %
        %
\begin{abstract}
        
We report the results of a new search for long range spin-dependent interactions using a Rb -$^{21}$Ne atomic comagnetometer and a rotatable electron spin source based on a SmCo$_{5}$ magnet with an iron flux return. By looking for signal correlations with the orientation of the spin source  we  set new constrains on the product of the pseudoscalar electron and neutron  couplings $g^e_p g^n_p/\hbar c<1.7\times10^{-14}$  and on the product of their axial couplings $g^e_A g^n_A/\hbar c<5\times10^{-42}$  to a  new particle with a mass of less than about $1~\mu$eV.   Our measurements improve by about 2 orders of magnitude previous constraints on such spin-dependent interactions. 
                        
\end{abstract}
\maketitle
        %

Long range interactions between spin-polarized  objects are  dominated by photon-mediated magnetic forces.  Additional long range forces may exist if there are  new light or massless particles beyond the Standard Model. For example, such new forces arise  from exchange of pseudoscalar  axions or axion-like particles ~\cite{moody1984new},  from spin-1 paraphotons or  light $Z'$ bosons ~\cite{dobrescu2006spin,safronova2018search}; from exchange of ``unparticles" \cite{Georgi2007,Liao2007}, dynamical breaking of local Lorentz invariance \cite{Arkani-Hamed2005}, or  propagating torsion in modified gravity \cite{Hammond1995,Naville1982}. In many of these models significant long-range interactions  appear only when both objects are spin-polarized, for example  for axion-like particles without a  $\bar{\theta }$ term \cite{moody1984new}, or for paraphotons-- massless gauge boson with dimension-6 operator coupling to fermions \cite{dobrescu2006spin}. Overall, search for axion or axion-like particle is of particular interest since they are candidates to explain the unexpected small level of CP violation in QCD or the nature of dark matter.

Experimental searches for  anomalous spin-spin interactions were first discussed by Ramsey \cite{ramsey1979tensor} and have been performed using  a variety of systems, including atomic comagnetometers \cite{aleksandrov1983restriction,vasilakis2009limits},  trapped ions  \cite{wineland1991search,kotler2015constraints}, spin-polarized pendulums \cite{heckel2013limits,terrano2015short}, polarized geoelectrons \cite{hunter2013using}, and NMR spectroscopy \cite{ansel1985restrictions,ledbetter2013constraints}. Such experiments typically use a ``spin source"-- a large collection of spin-polarized fermions
and a ``spin sensor"-- a sensitive system for measurement of the resulting shifts in spin energy levels.  Nuclear spin sensors typically have  good  energy resolution due to long spin coherence times of nuclear spin ensembles. Therefore, it is natural to combine a nuclear spin  detector, similar to the one used \cite{,vasilakis2009limits}, with a permanent magnet spin source that provides the highest density of polarized electron spins, as used in \cite{heckel2013limits,terrano2015short}.

Here we describe such an experiment searching for  electron-nuclear spin-dependent forces using a rotatable SmCo$_5$ spin source \cite{Heckel2006} and a $^{21}$Ne-Rb comagnetometer \cite{smiciklas2011new}. SmCo$_5$ has a unique property that part of its magnetization is created by angular moment of the electrons, instead of their spins. This allows one to cancel the net magnetic field created by the spin source without canceling an anomalous spin-dependent force.  Our experimental arrangement is sensitive to two spin-dependent potentials in the notation of  \cite{dobrescu2006spin} given  by:
\begin{eqnarray}
V_2=&&\frac{g^{1}_{A} g^{2}_{A}}{4 \pi}\frac{(\hat{\sigma}_1.\hat{\sigma}_2)}{r}e^{-r/\lambda}\label{eq:boson-1}\\ 
V_3=&&\frac{g^{1}_{p} g^{2}_{p}\hbar^{2}}{16 \pi M_1 M_2 c^2}\bigg[(\hat{\sigma}_1.\hat{\sigma}_2)\big(\frac{1}{\lambda r^2}+\frac{1}{r^3}\big)- \nonumber  \\ 
&&(\hat{\sigma}_1.\hat{r})(\hat{\sigma}_2.\hat{r})\big(\frac{1}{\lambda ^{2}r}+\frac{3}{\lambda r^2}+\frac{3}{r^3}\big)\bigg] e^{-r/\lambda} \label{eq:boson-0}
\end{eqnarray}
In the above equations, $\hat{\sigma}_i$ is the normalized expectation value of the $ i^{\rm th}$ particle spin and $M_i$ is its mass, $\lambda =\hbar /mc $ is the Yukawa range of the new particle with  mass $m$  mediating the spin-dependent force, and $r$ is the distance between the two  spins.   We set  new limits on the product of electron and neutron  pseudoscalar coupling constants $g^e_p g^{n}_p$ for $V_3$ and the product of the axial vector coupling constants $g^e_A g^{n}_A$ for $V_2$. The interaction potential $V_3$ can also be generated by a vector particle, such as a paraphoton or $Z'$ boson. Our measurements set new limits on the combinations of their parameters described  \cite{dobrescu2006spin}. One can also set limits in the product of electron and proton spin couplings using the sub-leading proton spin polarization in $^{21}$Ne \cite{brown2017nuclear}. Our limits are substantially better than can be extracted by combining the results of previous electron-electron and nuclear-nuclear spin force experiments.

 The Rb-$^{21}$Ne comagnetometer used in this experiment is similar to the one in \cite{smiciklas2011new}. More detailed explanation of an its operation can be found in \cite{smiciklas2011new,brown2010new,kornack2005nuclear,Brownthesis}. At the heart of the comagnetometer is an aluminosilicate GE180 spherical glass-blown cell $1\,$cm in diameter containing $1.5$ amagats of $^{21}$Ne, $50$ Torr of N$_{2}$ (to prevent radiation trapping), $^{87}$Rb and trace amounts of Cs. The cell is heated up to $180\,^{\circ}$C to create a dense, optically thick vapor of $^{87}$Rb. The Cs vapor remains optically thin and is optically pumped to create a relatively uniform spin polarization, which is transferred by spin-exchange collisions to Rb and then to $ ^{21}$Ne~\cite{romalis2010hybrid}. Cs is optically pumped using 450 mW of circularly polarized light at 895 nm.  

The spin polarization of $^{87}$Rb is measured via Faraday rotation of a linearly polarized probe beam  detuned from $795\,$nm and propagating through the cell in the $\hat{x}$ direction. To measure small optical rotation   the linear polarization of the probe beam is modulated at $50\,$kHz by a photoelastic modulator (PEM) and readout using a lock-in amplifier. Low frequency noise from air currents is greatly reduced by operating the experiment inside a vacuum bell jar at pressure of about $1\,$Torr. The probe and pump beams are  steered to illuminate the center of the cell, which reduces any spurious effects due to the linear dichroism of the cell walls~\cite{kornack2005test}.

\begin{figure} [ht]
\includegraphics[width=86mm]{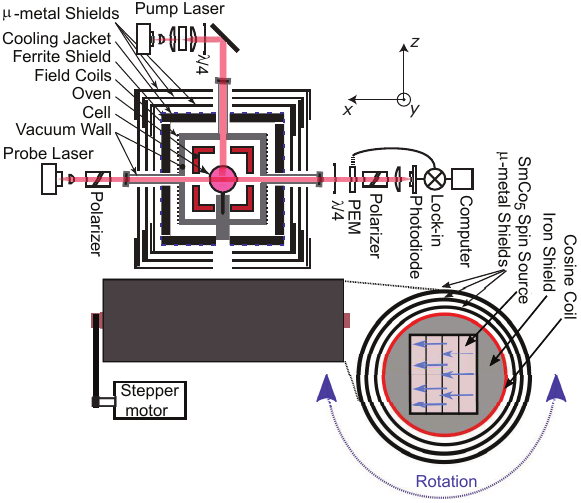}
\caption{The experimental setup, adapted from~\cite{brown2010new}. The spin source is placed under the comagnetometer cell with the rotation axis parallel to the probe beam.\label{fig:sketch}}
\end{figure}

The comagnetometer is operated at a compensation point where the external $B_z$ field is equal and opposite to the sum of the  effective magnetic fields created due to spin-exchange collisions with polarized   $^{87}$Rb and $^{21}$Ne \cite{kornack2005test}. Automated zeroing routines are used to adjust  the magnetic fields inside the shields in the $x$, $y$ and $z$ directions.  After field zeroing the leading term in the comagnetometer's signal at the compensation point is:
\begin{equation}
        S=\kappa\frac{\gamma_e P^e_z}{R_{tot}}\bigg[b_y^n-b^e_y+\frac{\Omega_y}{\gamma_n}\bigg],
\end{equation}where $\gamma_e$ and $\gamma_n$ are the gyromagnetic ratios of the free electron and $^{21}$Ne, respectively, $P^e_z$ is the polarization of  $^{87}$Rb, $R_{tot}$ is the total relaxation rate of $^{87}$Rb, and $b^{n,e}_y$ are the anomalous magnetic fields that couple to the $^{21}$Ne  and $^{87}$Rb electron spins in the $y$ direction. $\Omega_y$ is the gyroscopic rotation around the y-axis. The comagnetometer has suppressed sensitivity to ordinary magnetic fields but  retains leading order sensitivity to anomalous fields.  To calibrate the comagnetometer we measure the response to a slowly modulated $B_x$ field ~\cite{Brownthesis}. We verify the calibration factor by measuring the gyroscopic signal due to  a slowly changing tilt of the optical table and compare it to  the response of the tilt sensors.

The spin source is made from multiple rectangular blocks of  SmCo$_5$ permanent magnet with $7.6\times7.6\times20.3\,$cm$^3$ total volume surrounded by a cylindrical soft iron flux return with an outer diameter of 15.2 cm and length of 22.9 cm. The axis of the spin source is  25 cm away from the center of the comagnetometer cell. The remnant magnetic field outside of the iron cylinder is about $0.6\,$ mT, in good agreement with   finite element magnetic field analysis. To further   reduce this field, a cosine coil is mounted on the outside of the iron cylinder and three layers of $\mu$-metal shields are added around the spin source. The coil allows us to cancel the residual leakage of the fields by a factor of 10 or alternatively increase the field to check for systematic sensitivity of the comagnetometer. The orientation of the spin source is reversed by rotating the cylinder around its axis using a stepper motor and a timing belt while keeping the outer two magnetic shield layers  fixed. The residual magnetic field correlated to the orientation of the spin source is equal to approximately  $2.5\times10^{-9}\,$T. The comagnetometer apparatus is vibrationally isolated from the mechanical rotation system inside the vacuum bell jar.  

Fully magnetized SmCo$_5$ with $B_0 \approx 1$ T has an electron spin density of $4.5\times\, 10^{22}\,\text{spin}/\text{cm}^3$~while soft iron with the same magnetization has a spin density of $8.2\times\, 10^{22}\,\text{spin}/\text{cm}^3$~ \cite{heckel2008preferred,ji2017searching}. Hence the spin source posses a large net electron spin while having only a small residual magnetic dipole moment. The presence of net spin in a similar structure had been verified in \cite{heckel2008preferred} by observing the gyrocompass signal. The use of magnetic shielding does not screen anomalous spin interactions. Magnetic shielding around the spin source has a similar spin content to the soft-iron flux return. The magnetic shielding around the comagnetometer cannot hide the signal, as discussed in~\cite{kimball2016magnetic},  because we compare spin interactions of electron and nuclear spins in the $^{87}$Rb-Ne comagnetometer.  The rotation of the spin source is controlled by a separate computer to minimize possible cross-talk with the main system operating the experiment.

 Data is collected in intervals of $250\,$s  during which  the spin source is rotated by 180$^{\circ}$ 19 times every 12 seconds.   String analysis \cite{kornack2005test} is used to calculate the correlation of the comagnetometer signal with spin source orientation, using only the last 2 seconds to allow the system to settle mechanically after each rotation. The $B_z$ field is adjusted  at the end of each interval, while the other field components are zeroed and the comagnetometer sensitivity is calibrated every seven hours. 
 \begin{figure}
        \centering
        \includegraphics[width=80mm]{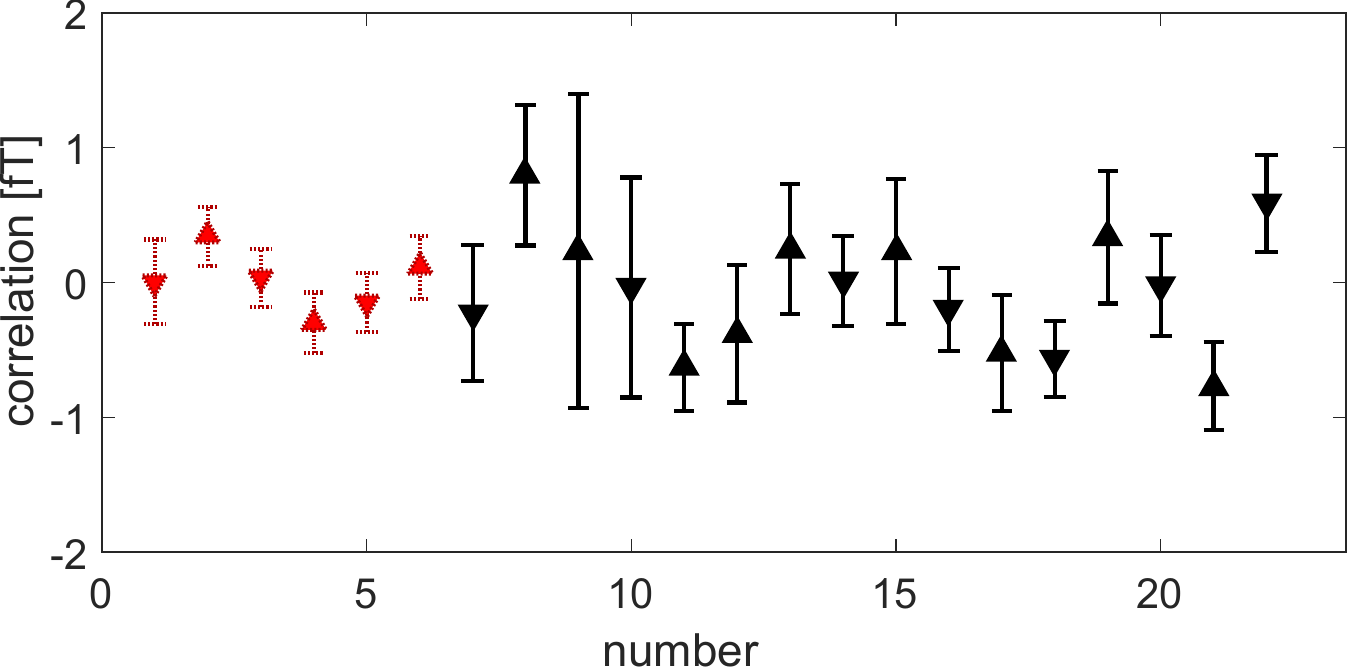}
        \caption{Measured correlation for positive polarization (red), negative polarization (black), clockwise spin source rotation (triangles pointing down), counterclockwise ones (triangles pointing up).\label{fig:raw}}
\end{figure}

\begin{table}
        \setlength{\tabcolsep}{7pt}
        \centering
        \begin{tabular}{ccc}
                \hline\hline
                Type & Weighted averaged correlation (aT)   & Reduced $\chi^2$ \\ 
                \hline
                $b^y_{p-}$              & $-180\pm110$   & 1.53           \\  
                $b^y_{p+}$              & $-9\pm83$   & 0.743             \\
                $b^y_{ccw}$             & $-71\pm89$ &      1.07       \\
                $b^y_{cw}$             & $-140\pm140$  & 1.97             \\
                \hline
                $b^y_{tot}$            & $-80\pm70$ & 1.32            \\
                \hline\hline
        \end{tabular}
        \caption{Measured correlation for negative polarization ($b^y_{p-}$), positive polarization ($b^y_{p+}$), counter clockwise spin source rotation ($b^y_{ccw}$), clockwise spin source rotation ($b^y_{cw}$), and total correlation ($b^y_{tot}$).}
        \label{tab:corr}
\end{table}

\begin{table}
        \setlength{\tabcolsep}{7pt}
        \centering
        \begin{tabular}{ccc}
                \hline\hline
                Sensors &  Averaged correlation (aT)    \\ 
                \hline
                Fluxgate X             & $-41\pm1$             \\  
                Fluxgate Y             & $48\pm1$              \\
                Fluxgate Z             & $-52\pm1$             \\
                Probe Beam Position (H)         & $ -14\pm38$           \\
                Probe Beam Position (V)          & $ 12\pm11$             \\
                Probe Beam Power        & $ 8\pm10$              \\
                Pump Beam Position   (H)       & $ -6.1\pm16$          \\
                Pump Beam Position   (V)       & $ 91\pm138$           \\
                Pump Beam Power        & $ -31\pm26$            \\
                Tilt Rate Y ($\Omega_y$)     & $-110\pm82$           \\
                Tilt Rate X ($\Omega_x$)    & $-1.4\pm9$            \\
                \hline
                Total &$-96\pm169$ \\
                \hline\hline
        \end{tabular}
        \caption{Signal correlation estimated from external sensors.}
        \label{tab:other}
\end{table}     

Figure~\ref{fig:raw} shows the results of approximately two weeks of data. Each data point corresponds to a $\sim24$ hours measurement. We collected data with clockwise and counter-clockwise  spin source rotations and for two orientations of the atomic spin polarization. The results of the measurements
of $b_y^n-b^e_y$ are summarized in Table \ref{tab:corr}. The error bars are scaled by the value of reduced $\chi^2$. Extended discussion about the method used to obtain uncertainty and reduced $\chi^2$ can be found in~\cite{Brownthesis,lee2019new}.

To check for possible systematic effects correlated with spin source orientation we monitor the magnetic fields, tilts of the comagnetometer platform, positions of the laser beams, as well as other signals that did not show significant effects. Measurements of the magnetic fields at several positions around the apparatus with a fluxgate magnetometer have average correlated field amplitudes  of  $8.2\times10^{-10}\,$T, $2.4\times10^{-9}\,$T and $2.6\times10^{-10}\,$T for $\hat{x}$, $\hat{y}$ and $\hat{z}$ directions, respectively. The combination of magnetic shielding around the cell and the comagnetometer compensation  give an additional suppression of external fields by $5\times 10^{-8}$ for $\hat{x}$ and higher suppressions for  $\hat{y}$ and $\hat{z}$ axes. Two 4-quadrant photodiodes monitor the positions  and powers of the pump and probe beams. A separate set of measurements was used to find the correlation between the rotation of the spin source and the beams' positions for both clockwise and counterclockwise rotations of the spin source in the same analysis window as the main measurement. To estimate sensitivity to beam motion, larger beam motion was induced while monitoring the comagnetometer signal. A precision tiltmeter mounted on the same vibration-isolation platform as the comagnetometer measured the residual rotation rate of the platform correlated with spin source reversal. 

\tabref{tab:other} shows the summary of measured systematic effects. The total systematic error from magnetic field leakages, beam positions and power, as well as gyroscopic couplings is constrained with an uncertainty of $169\,$aT. Our assumption is that systematic effects recorded for the relevant sensor
correlations are independent. Hence, we can combine their uncertainties in quadrature to provide an estimate of the
overall systematic uncertainty. Hence, we quote the final total anomalous coupling as $b^y_{tot}=-80\pm 70_{stat}\pm169_{syst}$. This yields, at the $95\%$ confidence level, $|b^y_{tot}|<400$ aT. 

To convert the measured value of $b^y_n-b^y_e$ to  limits on  spin-spin interactions we  express the energy shift due to the anomalous potentials for neutrons, protons and electrons as $V_n f_n +V_p f_p -V_e =\mu_{^{21}Ne}b^n_{y}- \mu_B\b^{e}_{y}$, where $f_n=0.58$ and $f_p=0.04$ are the fraction of neutron and proton spin  polarization in $^{21}$Ne~\cite{brown2017nuclear}.  The interactions given by Eq. (\ref{eq:boson-1},\ref{eq:boson-0}) are integrated over the distribution of the polarized spins in the SmCo$_5$ magnets and the soft-iron flux return.   

\begin{figure} 
        \centering
        \includegraphics[width=86mm]{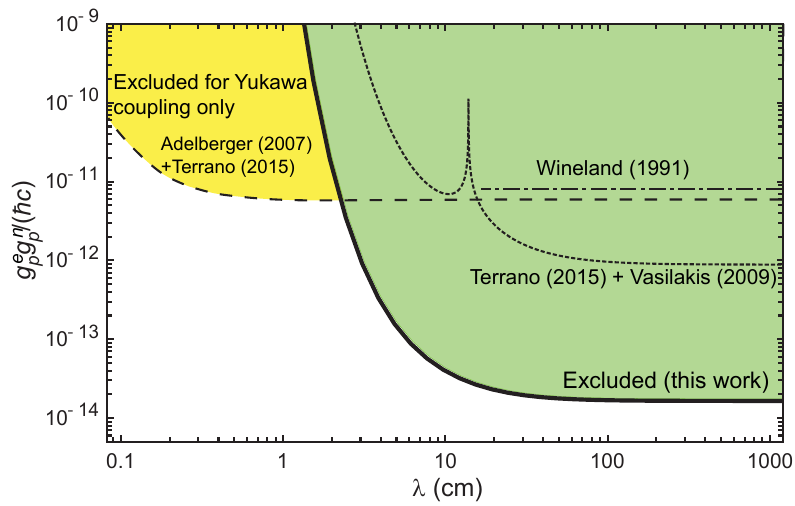}
        \caption{Constraints ($95\%$ CL) on the coupling parameter   $g^{n}_{p}g^{e}_{p}$ for two fermions interacting by a pseudoscalar boson.\label{fig:limitsv3}}
\end{figure}

\begin{figure}
        \centering
        \includegraphics[width=86mm]{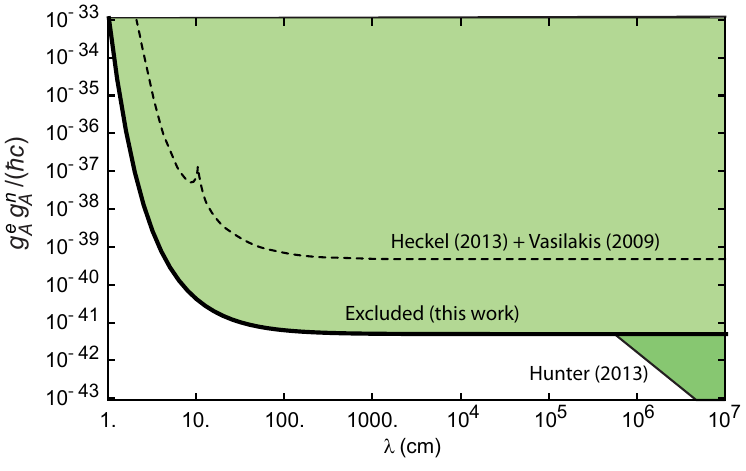}
        \caption{Constraints ($95\%$ CL) on the coupling parameters  $g^{n}_{A}g^{e}_{A}$  for two fermions interacting by a spin-1 boson.}\label{fig:limitsv2}
\end{figure}

The limits on the pseudoscalar and axial coupling constants are summarized in Table \ref{tab:limits} and Figs. \ref{fig:limitsv3}, \ref{fig:limitsv2}. Only one prior experiment has constrained directly the $g^{n}_{p}g^{e}_p$ combination \cite{wineland1991search}. More stringent limits can be obtained by combining the limit on $(g^{e}_p)^2$ from \cite{terrano2015short} and the limit on  $(g^{n}_p)^2$ from \cite{vasilakis2009limits}. If the pseudoscalar particle is coupled to fermions through a Yukawa interaction (as opposed to the derivative coupling typical for axions), one can also obtain a limit on $(g^{n}_p)^2$ from two-particle exchange using equivalence principle experiments \cite{Krauss}. Several additional limits can be set on combinations of coupling parameters for paraphoton and $Z'$ boson from the expressions for $V^{en}_3$ derived in \cite{dobrescu2006spin}. We also set limits on the product of axial couplings $g^{n}_{A}g^{e}_A$ for a vector boson exchange, improving on previous direct \cite{hunter2013using} and indirect limits \cite{heckel2013limits,vasilakis2009limits} for a particle with Yukawa range of $1 $ cm to $10^6$ cm. Several additional constraints on the order of $10^{-10}-10^{-18} $ exist for  $(g_{A})^2/\hbar c$  that extend to much shorter length scales \cite{ledbetter2013constraints,kotler2015constraints,ficek2018}.  


\begin{table}
\setlength{\tabcolsep}{3pt}
 {
  \begin{tabular}{|l|c|c|l|}
 \hline
Coupling    & This work   & Previous limit & Reference\\ 
\hline
$g^{e}_{p}g^{n}_{p}/\hbar c$ & $1.7\times10^{-14}$   & $8.1\times10^{-12}$&  Direct:~\cite{wineland1991search}\\
 & & $9.0\times10^{-13}$& $(g^e_p)^2(g^n_p)^2:$~\cite{terrano2015short,vasilakis2009limits}\\
 & &$5.9\times10^{-12}$& Only for Yukawa\\  & & & coupling:  ~\cite{terrano2015short,adelberger2007particle}\\
  \hline
 $g^{e}_{p}g^{e}_{p}/\hbar c$ & $1.5\times10^{-14}$  & $5.5\times10^{-17}$  &Direct:~\cite{ terrano2015short}      \\
 \hline
 $g^{e}_{A}g^{n}_{A}/\hbar c$ &  $5.0\times10^{-42}$  & $4.8\times10^{-40}$  & $(g^e_A)^2(g^n_A)^2$:~\cite{heckel2013limits,vasilakis2009limits}\\ 
 \hline
 $g^{e}_{A}g^{e}_{A}/\hbar c$             &  $8.0\times10^{-39}$  & $7.6\times10^{-40}$        &   Direct:~\cite{ heckel2013limits} \\ 
 \hline
 \end{tabular}}

 \caption{Experimental limits (95\% CL) on anomalous spin-spin interaction between two fermions by spin-$0$ or spin-$1$ boson exchange with  Yukawa range of $10^2$ to $10^6$ cm.}
                \label{tab:limits}
        \end{table}

In conclusion, we have improved limits on spin-dependent interactions between electrons and neutrons mediated by a new  light pseudoscalar or vector boson by about 2 orders of magnitude. The experimental uncertainties are dominated by mechanical transients which produce the largest systematic errors and force us to use a  short integration time and a slow source reversal. The sensitivity can be improved by about 2 orders of magnitude with a better vibration isolation system to reduce the  motion associated with the mechanical reversal of the spin source.  We like to thank Justin Brown, Lawrence Cheuk,  David Hoyos, and Ahmed Akhtar for assistance in the design and assembly of the spin source. This work was supported by NSF Grant No. 1404325.
        

\end{document}